# Parallel Magnetic Field Induced Glass-Like Behavior of Disordered Ultrathin Films

L. M. Hernandez, A. Bhattacharya, K. A. Parendo and A. M. Goldman

*School of Physics and Astronomy, University of Minnesota, 116 Church St. SE, Minneapolis, MN 55455, USA*

(November 21, 2018)

Dramatic glass-like behavior involving nonexponential relaxation of in-plane electrical conduction, has been induced in quench-condensed ultrathin amorphous Bi films by the application of a parallel magnetic field. The effect is found over a narrow range of film thicknesses well on the insulating side of the thickness-tuned superconductor-insulator transition. The simplest explanation of this field-induced glass-like behavior is a Pauli principle blocking of hopping transport between singly occupied states when electrons are polarized by the magnetic field.

Nonexponential relaxation effects extending over very long times can result from competition between interactions and disorder in systems that are glasses. Such effects have been found in strongly disordered electronic systems such as doped semiconductors and disordered metals that are described as Coulomb glasses, in particular in configurations of films grown on insulating substrates which isolate them from a gate electrode [1–4]. In these studies the response of the conductance of the film to capacitive charging using a gate exhibited hysteresis, slow nonexponential relaxation, and memory effects, behaviors characteristic of glasses. Such very slow time development of the response of Coulomb glasses in this geometry has been found in theoretical modeling by Yu [5] based on the model of Baranovskii, Shklovskii, and Efros [6].

In this letter we report very slow, glass-like behavior of the in-plane conductance of quench-condensed ultrathin films of amorphous Bi ($a$-Bi) upon application of a parallel magnetic field. The effect is found over a narrow range of thicknesses, or film sheet resistances, well on the insulating side of the thickness-tuned superconductor-insulator(SI) transition [7]. This behavior becomes more pronounced with increasing magnetic field, and disappears at temperatures above 200mK. The scenario in which a very slow glassy state involving *in-plane* conductance is induced by the application of a parallel magnetic field appears to be new and unique. Within the theoretical literature on glassy dynamics of disordered electronic systems, the most likely explanation would appear to be that proposed some years ago by Kurobe and Kamimura [8] and elaborated upon by Matveev *et al.* [9]. In this picture, two electrons with parallel spin cannot be localized within a single orbital. Thus when a magnetic field polarizes the spins, transfers of electrons between singly occupied states are blocked, and the rate of relaxation is greatly reduced.

Amorphous Bi films were grown using an apparatus that consisted of a specially prepared Oxford Instruments dilution refrigerator (Kelvinox 400) with a "bottom loading" sample transfer system, all maintained under ultra-high vacuum conditions. The access chamber served as a deposition chamber, with film growth being carried out while the substrate was attached to the liquid helium cooled loading stick. This apparatus can be used to study the evolution with thickness of the properties of films, which would anneal if warmed up. Low temperature measurements down to 0.050K and in magnetic fields of up to 12T were carried out with the sample attached to the mixing chamber of the refrigerator, and disconnected from the sample loading stick. Details of this setup are described elsewhere [10].. Although the ranges of magnetic field and temperature were substantially greater than those available in an apparatus used for earlier work [11], a nearly identical geometry of evaporation sources and ultra-high vacuum environment was provided. The vapor sources were commercial Knudsen cells, and the flux density at the substrate was uniform to better than one part in $10^4$, permitting effects associated with very tiny changes in thickness to be observed. The epi-polished $SrTiO_3$ wafers employed as substrates were first cooled to helium temperatures and then pre-coated with a thin layer of $a$-Ge, before the Bi growth was initiated. Films produced in this manner are believed to be disordered on microscopic length scales [12].

Four-terminal measurements were carried out using a Keithley model 220 current source and a Keithley model 182 nanovoltmeter. All electrical lines to the sample were filtered with a cutoff frequency of about 500Hz. The Keithley model 220 current source was filtered with a cutoff frequency of about 10Hz. The measuring system was optically isolated from the controlling computer.

Figure 1 shows $R(T)$ for a series of Bi films of different thicknesses. Films for which glassy behavior in a parallel magnetic field was observed, are shown in bold. As can be seen, all films were insulating at the very lowest temperatures, until the nominal film thickness was 13.35Å, at which point evidence of superconducting order was observed in the form of a drop in resistance between 200 and 400mK.

In the case of the four thinnest films, at temperatures above about 100mK, $R(T)$ could be fit by the functional form $\exp\left[-\left(\frac{T_0}{T}\right)^{0.66}\right]$. The value of the exponent, 0.66, is slightly different from 0.5, which is predicted by variable range hopping theory including Coulomb effects [13], but is consistent with many other measurements on disordered systems [14]. The magnetoresistive behaviors of these films were consistent with the theory of Efros and Shklovskii [13]. No glass-like behavior in a magnetic field was recorded in the case of the three thinnest films, because the phenomenon was only discovered during the study of the fourth film. The hopping exponents of films that did not exhibit glass-like behavior, but which were nevertheless insulating, decreased with increasing film thickness, possibly as a consequence of contributions from conductance channels associated with superconductivity or superconducting fluctuations, which as mentioned above, first appeared in a 13.35Å thick film.

The glass-like behavior found in parallel magnetic fields manifests itself in a four-terminal measurement as a very slow nonexponential buildup of the current through the film accompanied by memory effects. The method used to make the measurements involved first setting the current flowing in plane through the film using a constant current source. In zero magnetic field, voltage, as measured across the voltage leads, developed in response to this current in a time the order of the total RC time constant of the circuit. In a magnetic field, the 10V voltage compliance limit of the current source was reached in a time somewhat longer than the time-constant of the circuit, and the voltage across the current leads was then clamped at that value. The current source is then acting as a constant voltage source. The current through the film continued to build very slowly, as evidenced by an increasing voltage across the voltage leads that changes in a nonexponential manner over times that are orders of magnitude longer than the circuit time constant.

Glass-like behavior was found in fields ranging from 0.1 T up to 12 T. The change of conductance with time was slower the higher the magnetic field. Presumably increasing the magnetic field increases the degree of spin polarization of the carriers. Figure 2 shows the voltage response of an 11.38Å thick film in both zero field and in a field of 12T. In zero magnetic field there is essentially no change in the voltage drop over very long times. On the other hand, in a 12T field, the voltage grows at long times as $lnt$, where $t$ is time. The fraction of the measured voltage drop exhibiting glass-like behavior contributes about 1/3 of the measured conductance in a field of 12T and a temperature of 50mK, after about $10^4$ seconds. The drift curves studied in some intermediate magnetic fields did not exhibit simple logarithmic dependence on time, and in some instances did not change monotonically with time. Variations with a simple logarithmic form were found only in the largest fields. When the temporal evolution was not monotonic, the voltage output of the current source itself was changing, in response to the effective resistance of the film falling enough to unclamp the voltage of the current source. This was considered to be an artifact of the measurement technique and for this reason is not shown.

The thickness dependence of the glass-like behavior at a fixed magnetic field of 12T and at a fixed temperature of 50mK are shown in Fig.3. The fraction of the conductance responding in a glass-like manner decreased with increasing film thickness and the effect vanished completely when the film thickness reached 11.95Å. Glass-like behavior was never observed at temperatures above 200mK. The extraordinarily slow response precluded any truly systematic examination of the dependence on thickness, temperature, and magnetic field. As a consequence, it is not known whether there is a well-defined glass transition as a function of any of these variables.

Other evidence of glass-like behavior was found in the form of memory effects. One example of this is the following experiment: with the temperature at 50mK, and the field set at 12T, and the current source connected, the increase in the voltage drop across the sample was monitored. The current source was then switched off, and the voltage measured was observed to fall to a low value in a time given by the circuit time constant. Then the current source was restored, and the voltage was observed to rise within the circuit time constant to the value it exhibited before the current source was disconnected, and then to resume its very slow drift. This is shown in Fig. 4a. In Fig. 4b, we show another memory effect. Starting at a temperature of 50mK, the current source was turned off and the temperature raised to 100mK. The current was then restored and the voltage was seen to decay down to a new value, at which point the upward conductance drift resumed, at a rate governed by the new temperature.

We suggest that these effects are not due to a perpendicular component of magnetic field resulting from inaccuracy in positioning the plane of the film relative to the field direction, because of the smallness of this inaccuracy. The latter we estimate to be less than 1 degree, or 1/60 of a radian. In the highest and lowest fields in which glass-like phenomena were seen, 12 T and 0.1 T, this would result in perpendicular components of magnetic field of less than 0.2 T and 0.0015T respectively. We were concerned that these effects might be a consequence of drift in the magnetic field produced by the superconducting magnet, which was operated in the persistent current mode. We repeated the measurements with the magnet in the driven mode and found identical results.

Most experimental studies of the response of insulating two dimensional electron systems to parallel magnetic fields, other than the work of Ovadyahu and Pollak [2] and Yu [5], have been focussed on magnetoresistive behavior. Mertes and co-workers [15] studied dilute insulating two-dimensional electron gases in Si MOSFET configurations, and found large initial increases in resistivity with increasing field, followed by a saturation of the resistance with further increase in field. This was unexpected behavior for systems exhibiting variable range hopping, and is a larger effect than the positive magnetoresistance that can result if alignment of electron spins by a parallel field suppresses hops between singly occupied states because of the exclusion principle [8]. The latter idea has been used to explain a small positive magnetoresistance in $In_2O_3$ films [16], and is believed to be the basis, as discussed below, for the very slow glass-like behavior reported here. Parenthetically, in our films a very small magnetoresistance of order of a few percent was found in fields ranging from 0.03 T up to 12T [17], but was impossible to quantify because of the extremely slow response of the film. We did not see evidence of a magnetic field-induced quantum metal phase reported recently by Butko and Adams for amorphous Be films [18]. The fact that glass-like effects appeared in a 0.1T field and disappeared completely above 200mK, suggests that the field of onset, and the temperature of disappearance of the effect scale together approximately as $\mu_B H \approx k_B T$.

The simplest explanation of the phenomenon we have found derives from the above correspondence. With disorder in a certain range, the energy level spacings in the system and the localization radius of the electronic states are such that the application of a parallel magnetic field introduces a critical additional constraint, shutting off hopping transitions between singly occupied states, thus reducing the conductivity and dramatically increasing the time of relaxation. The effect becomes more pronounced the larger the magnetic field, and disappears when the disorder drops below some threshold. The mechanism would be relevant only if a fraction of sites can accommodate more than one electron. As pointed out by Matveev et al [9], this can occur if the on-site Coulomb repulsion U between electrons is smaller than the width of the distribution function of the energies of the localized states. There are then two types of sites that can contribute to hopping transport, those which have energies close to the Fermi level $\mu$ (Type I) and those which have one electron at a deep level with an energy of order $\mu - U$ (Type II). Other types of sites will have energy levels too far from the Fermi level to contribute to transport. In zero magnetic field the probability for two electrons on two singly occupied sites to have opposite spins is 1/2. As a consequence, transitions between Type I and Type II sites are possible. In a sufficiently strong magnetic field, these hops would be completely suppressed if one assumed that two electrons on the same site form a singlet state in the field. The effect would disappear in thicker films because the role of hopping would be diminished. It is not clear from these arguments as to why the drift should be in the direction of higher conductivity. Although the above may be the simplest explanation of our observations, it does not provide a quantitative picture of the data.

We acknowledge very useful discussions with F. Zhou and L. I. Glazman. This work was supported by the National Science Foundation under grant # NSF/DMR-987681.

FIG. 1. Sheet resistance vs. temperature for a series of films with thicknesses, from top to bottom, of 11.15, 11.25, 11.37, 11.38, 11.43, 11.48, 11.55, 11.65, 11.75, 11.85, 11.95, 12.03, 12.17, 12.27, 12.4, 12.55, 12.65, 12.85, and 13.35 Å. Those films that exhibited glass-like behavior as described in the text are in bold.

FIG. 2. Voltage vs. the natural logarithm of time for an 11.38Å thick film in zero magnetic field and in a field of 12T.

FIG. 3. Voltage vs.time for a series of films of different thicknesses at a temperature of 50mK and in a magnetic field of 12T. From top to bottom they are 11.38, 11.48, 11.55, 11.65, 11.75, and 11.85Å. The vertical axis has been adjusted to take out contributions from small voltage offsets that are due to thermal emfs that cannot be removed in the standard way as the bias must remain fixed. The data for the for the films with the two greatest thicknesses overlap.

FIG. 4. (a) Voltage vs time for an 11.65 Å thick film in a parallel field of 12T. At approximately 8500 seconds into the drift the current source was disconnected for about $10^3$ seconds, and then reconnected. Thedrift resumed from a value close to that achieved when the source was intially disconnnected. (b) Drift of the conductance of an 11.65Å thick film in a field of 12 T, initially at 50mK. The source was disconnnected, and the sample warmed to 100mK. The curve shows the variation of the voltage with time after this was done.